\documentclass[fp,twocolumn]{jpsj3}
\usepackage{txfonts}
\usepackage{color}
\usepackage{upgreek}

\definecolor{newgreen}{rgb}{0.203,0.531,0.234}

\newcommand{\VFe}{Sr$_2$VFeAsO$_3$}
\newcommand{\VFed}{Sr$_2$VFeAsO$_{3-\delta}$}
\newcommand{\Tc}{$T_{\mathrm{c}}$}
\newcommand{\Tstar}{$T^{\ast}$}
\newcommand{\Tsstar}{$T^{\ast\ast}$}

\title{Role of Vanadium-Oxide Layer in Electronic State of \VFed{} \\ with Oxygen Deficiency}

\author{Masamichi Nakajima$^1$\thanks{Present address: RIKEN Center for Emergent Matter Science, Wako, Saitama 351-0198, Japan \newline E-mail: masamichi.nakajima@riken.jp}, Hiroaki Yokota$^1$, Taihei Wakimura$^1$, Tetsuya Takeuchi$^2$, Koya Nakamura$^3$, Mitsuharu Yashima$^3$, Hidekazu Mukuda$^3$, Shigeki Miyasaka$^1$, and Setsuko Tajima$^1$}
\inst{$^1$Department of Physics, Osaka University, Toyonaka, Osaka 560-0043, Japan \\
$^2$Cryogenics Support Division, Core Facility Center, Osaka University, Toyonaka, Osaka 560-0043, Japan \\
$^3$Graduate School of Engineering Science, Osaka University, Toyonaka, Osaka 560-8531, Japan} 

\abst{Iron-based superconductor \VFe{} is composed of alternate stacking of a superconducting FeAs layer and an insulating vanadium-oxide layer with a perovskite-type structure. Electronic orders stemming from the spin and orbital degrees of freedom of V $3d$ electrons can arise in the vanadium-oxide layer, but such orders have not been confirmed so far. Here, we systematically investigate the electronic state of \VFed{} with oxygen deficiency and demonstrate the phase diagram of \VFed{} as a function of the $c$-axis lattice parameter, which has turned out to be a suitable measure of the amount of oxygen deficiency. We found a magnetic and structural anomaly at $\sim 100$ K with a thermal hysteresis, which is manifested with the introduction of oxygen deficiency. The presence of orthorhombic distortion was revealed below the temperature at which the anomaly appears, suggestive of V orbital ordering involving the $d_{xz}$ and $d_{yz}$ orbitals. It seems that substantial fluctuations associated with the orthorhombic distortion significantly influence the electronic state of the FeAs layer. Our findings indicate that the vanadium-oxide layer plays a significant role in the electronic state of \VFed{}.}

\begin{document}
\maketitle

\section{Introduction}

Iron-based superconductors are layered compounds, which consist of alternate stacking of superconducting FeAs layers and other spacer layers. The diversity of spacer layers gives rise to the variety of the families of iron-based superconductors~\cite{Johrendt2011,Jiang2013,Luo2015}. Introduction of chemical substitution or oxygen deficiency into spacer layers can control carrier density, which is one of the effective control parameters to achieve superconductivity, with keeping the FeAs layers intact. The coexistence of superconductivity and magnetism has been reported in iron-based superconductors with the spacer layer containing Eu~\cite{Zapf2017}. Various peculiar phenomena, such as a spontaneous vortex state~\cite{Jiao2017,Stolyarov2018}, an unusual enhancement of magnetic induction~\cite{Vlasko-Vlasov2019}, and a ferromagnetic order induced by superconducting vortices~\cite{Ishida2021}, have been observed due to the interplay between superconductivity and magnetism. Not only magnetism, but also other electronic orders originating from the spacer layer could correlate with superconductivity, which would lead to novel quantum phenomena.

\begin{figure}
\includegraphics[scale=0.98]{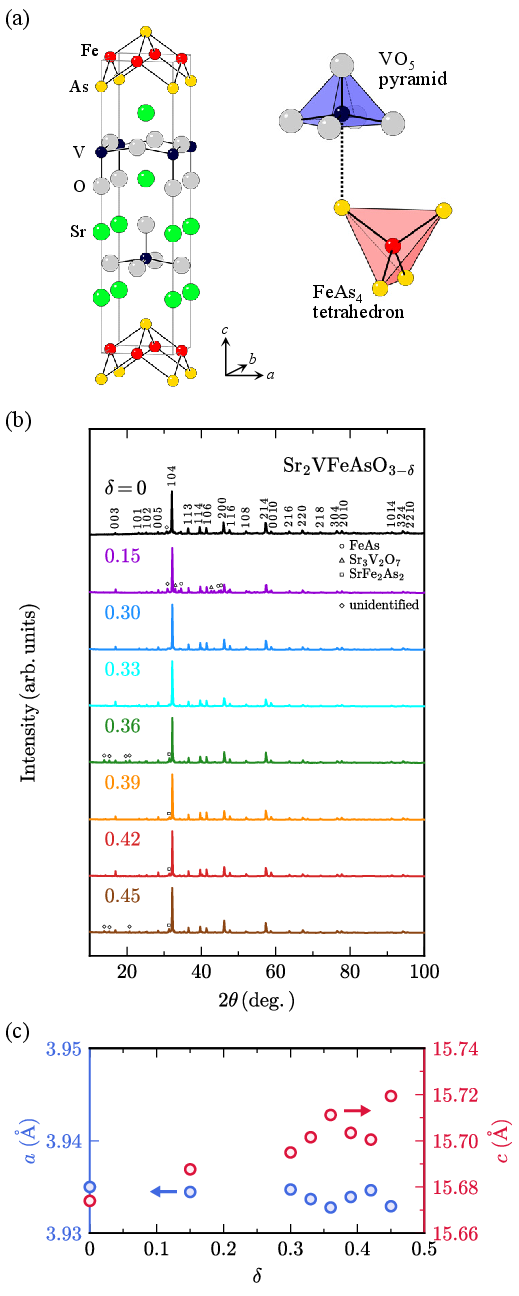}
\caption{(Color online) (a) Crystal structure of \VFe{}. The local structure of a VO$_5$ pyramid and an FeAs$_4$ tetrahedron is displayed at the right. The As atoms are located just below (or above) the V atoms. (b) Powder XRD pattern of \VFed{} with various values of $\delta$ measured at room temperature. Open symbols indicate peaks from impurity phases. (c) Variation of the $a$- and $c$-axis lattice parameters obtained from the XRD patterns as a function of $\delta$.}
\label{XRD}
\end{figure}

\VFe{} is a member of iron-based superconductors with the superconducting transition temperature \Tc{} exceeding 30 K, characterized by a thick spacer layer with a perovskite-type structure [Fig.~\ref{XRD}(a)]~\cite{Zhu2009}. The structure of the spacer layer is similar to the K$_2$NiF$_4$-type structure, which is well-known as a layered perovskite-type structure, and the lack of one O atom leads to a formation of a VO$_5$ pyramid. Band calculations initially suggested that V electrons contribute to the construction of the Fermi surface of \VFe{}~\cite{Shein2009,Wang2009,Mazin2010,Lee2010}. However, the Fermi surface similar to other iron-based superconductors has been experimentally confirmed~\cite{Qian2011,Kim2015}. This indicates that the V $3d^{2}$ electrons are localized due to strong electronic correlations~\cite{Nakamura2010}, while the Fe $3d^{6}$ electrons are itinerant responsible for superconductivity. The nominally realized V $3d^2$ state means that both the spin and orbital degrees of freedom are active. Vanadium oxides with the K$_2$NiF$_4$-type structure, like Sr$_2$VO$_4$~\cite{Sakurai2015,Teyssier2016} and LaSrVO$_4$~\cite{Dun2014}, exhibit novel electronic states involving the spin and orbital degrees of freedom. Such a behavior can arise in the vanadium-oxide layers of \VFe{}, which would lead to a nontrivial electronic phase via an interplay entangled with superconductivity. A notable feature of \VFe{} is the presence of a second-order phase transition at \Tstar{} $\simeq 150$ K without showing magnetic ordering or breaking tetragonal lattice symmetry~\cite{Cao2010,Tatematsu2010,Hummel2013}. The origin of the phase transition remains to be elusive. A charge/orbital order in the FeAs layers has been proposed as an origin~\cite{Ok2017}, but the vanadium-oxide layer is also considered to play some role because this phase transition has not been observed for other iron-based superconductors.

The electronic state of \VFe{} can be controlled by introduction of oxygen deficiency. It has been reported that oxygen deficiency makes a significant influence on the physical properties~\cite{Han2010,Sefat2011,Che2014,Tojo2019}. \Tc{} tends to decrease with oxygen deficiency~\cite{Han2010}. The variation in reported values of \Tc~\cite{Zhu2009,Cao2010,Tegel2010,Munevar2011,Holenstein2021} is due to the difference in the amount of oxygen deficiency. A spontaneous magnetization has been observed for \VFed{} without showing superconductivity~\cite{Tojo2019,Sefat2011}. Elucidation of the effect of oxygen deficiency is indispensable for deeply understanding the electronic state of this system, but there are few studies that have systematically investigated the variation of the electronic state of \VFed{} with oxygen deficiency.

To clarify the contribution of the vanadium-oxide layer to the electronic state of \VFed{}, we focused on oxygen deficiency, which affects the physical properties of this system. In this paper, we synthesized polycrystalline samples of \VFed{} with various values of  nominal $\delta$ and systematically investigated the variation of the electronic state with oxygen deficiency. We demonstrated the phase diagram of \VFed{} as a function of the $c$-axis lattice parameter as a measure of the amount of oxygen deficiency. It was found that besides the phase transition at \Tstar{}, an anomalous behavior in both magnetic and structural properties appears at \Tsstar{} $\simeq 100$ K with a thermal hysteresis, which is pronounced for the samples with oxygen deficiency. This anomaly can be attributed to orbital ordering involving the $d_{xz}$ and $d_{yz}$ orbitals of V. Substantial orbital fluctuations associated with this orbital ordering seem to be present above \Tsstar{}, which is likely linked to the disappearance of the $^{75}$As-NMR signal for a wide temperature range between \Tstar{} and \Tsstar{}. The present results demonstrate that the insulating vanadium-oxide layer plays a significant role in the electronic state of \VFed{}.

\section{Experimental Methods}

Polycrystalline samples of \VFed{} were synthesized by a solid-state reaction of SrO, V$_2$O$_5$, Fe, V, and Sr$_3$As$_2$. SrO was prepared by heating SrCO$_3$ at 1320$^{\circ}$C for 30 h. Sr$_3$As$_2$ was prepared by heating Sr and As at 500$^{\circ}$C for 10 h and then at 650$^{\circ}$C for 10 h. The starting materials were mixed with a nominal composition of \VFed{} and pressed into a pellet in a glove box filled with high-purity argon gas. The pelletized mixture was sealed in an evacuated quartz tube, followed by a heat treatment at 900$^{\circ}$C for 20 h and then at 1150$^{\circ}$C for 20 h. The calcined pellet was again ground, pelletized, and sealed in a quartz tube, followed by heating at 1150$^{\circ}$C for 20 h.

Powder x-ray diffraction (XRD) patterns were measured at room temperature using a Rigaku RINT-2000 diffractometer with Cu K$\alpha$ radiation. Low-temperature XRD measurement was performed with the photon energy of 11.5 keV at BL-8A of Photon Factory (PF) in KEK, Japan. Magnetic susceptibility was measured using a Quantum Design magnetic property measurement system. Resistivity measurements were conducted by a standard four-terminal method. Specific heat was measured by the quasi-adiabatic heat-pulse method in the temperature range between 4.2 and 200 K using a homemade apparatus. $^{51}$V- and $^{75}$As-NMR measurements were carried out using a conventional pulsed NMR spectrometer in magnetic fields at a fixed frequency 77.11 MHz.

\section{Results}

\subsection{Variation of Lattice Parameters of \VFed{} with Oxygen Deficiency}

Figure \ref{XRD}(b) shows the XRD patterns of \VFed{} with various nominal values of $\delta$. The $a$- and $c$-axis lattice parameters derived from the XRD patterns are plotted in Fig.~\ref{XRD}(c). The $c$-axis lattice parameter tends to increase with the nominal amount of oxygen deficiency, consistent with the previous reports~\cite{Che2014,Tojo2019}. The elongation of the $c$-axis length with oxygen deficiency was also observed in Sr$_2$VO$_4$~\cite{Ueno2014}, which possesses the same layered perovskite-type vanadium-oxide layers as \VFed{}. The variation of $a$ with $\delta$ is small but shows a clear inverse correlation with $c$. The pronounced elongation of the $c$-axis length implies that oxygen vacancies are introduced into the SrO layer, consistent with the study of a transmission electron microscopy, which demonstrated that oxygen deficiency has a significant influence on the SrO layers~\cite{Che2014}. The oxygen vacancies in the SrO layer was also supported by the theoretical prediction in terms of total energies~\cite{Suetin2012}, although the theoretically calculated change in the lattice parameters is the opposite.

Since the actual amount of oxygen deficiency can deviate from nominal $\delta$, the $c$-axis lattice parameter, rather than $\delta$, can be regarded as a suitable measure. The variation of $c$ is very clear, and the physical properties of \VFed{} systematically change with $c$ as will be shown later. In the present study, the $c$-axis lattice parameter for $\delta=0$ was estimated to be 15.6740(12) \AA, comparable to the reported values for the samples with \Tc{} exceeding 30 K~\cite{Zhu2009,Tatematsu2010}. For $\delta=0.36$, the estimated value of $c$ is higher than that for $\delta=0.39$ and 0.42, indicating that the sample with $\delta=0.36$ contains more oxygen vacancies.

\subsection{Suppression of Superconductivity with Oxygen Deficiency}

\begin{figure}
\includegraphics[scale=0.98]{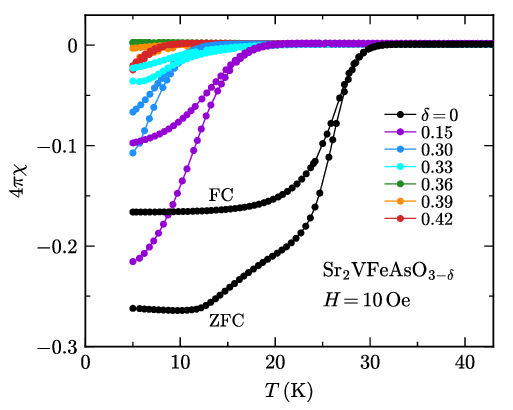}
\caption{(Color online) Temperature dependence of magnetic susceptibility of \VFed{} in zero-field-cooling (ZFC) and field-cooling (FC) conditions under $H = 10$ Oe.}
\label{chiSC}
\end{figure}

\begin{figure}
\includegraphics[scale=1.00]{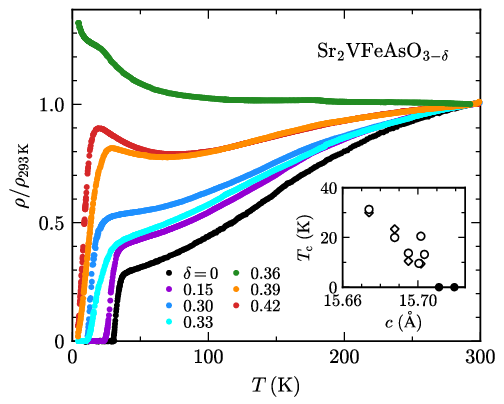}
\caption{(Color online) Temperature dependence of resistivity $\rho$ of \VFed{} normalized at 293 K. Inset shows the variation of \Tc{} as a function of the $c$-axis lattice parameter. The \Tc{} values were determined from the onset of magnetic susceptibility (open circles) and zero resistivity (open diamonds). The samples for which superconductivity was not observed in this study are indicated by closed circles.}
\label{resistivity}
\end{figure}

The introduction of oxygen deficiency significantly affects superconductivity in \VFed{}. Figure \ref{chiSC} shows the temperature dependence of magnetic susceptibility of \VFed{} under a magnetic field $H$ of 10 Oe. The onset temperature of a diamagnetic response for $\delta=0$ exceeds 30 K with the largest superconducting volume fraction. With increasing $\delta$, \Tc{} tends to decrease. Superconductivity was not observed down to 5 K for $\delta=0.36$ with largest $c$ among the samples presented in Fig.~\ref{chiSC}. Note that \Tc{} for $\delta=0.33$ is larger than that for $\delta=0.30$ and 0.42 with a shorter $c$ axis, suggesting that the suppression of \Tc{} may not be monotonic with respect to the amount of oxygen deficiency (see the inset of Fig.~\ref{resistivity}).

The suppression of superconductivity with oxygen deficiency was also confirmed by the temperature dependence of resistivity $\rho(T)$. As shown in Fig.~\ref{resistivity}, \Tc{} decreases with increasing $\delta$. The oxygen deficiency induces an appreciable reduction of a residual resistivity ratio, consistent with the previous studies~\cite{Han2010,Sefat2011}. We observed a low-temperature upturn of $\rho(T)$ for the samples with large $\delta$. These facts indicate that oxygen deficiency causes a rather strong disorder effect on the conduction electrons despite the introduction of oxygen vacancies presumably in the SrO layers farthest from the conducting FeAs layers. The variation of \Tc{} with oxygen deficiency is summarized in the inset of Fig.~\ref{resistivity} as a function of $c$, demonstrating the tendency for the suppression of superconductivity with oxygen deficiency.

\subsection{Two Anomalies of \VFed{} Manifested in Magnetic and Structural Properties}

\begin{figure*}
\includegraphics[scale=0.96]{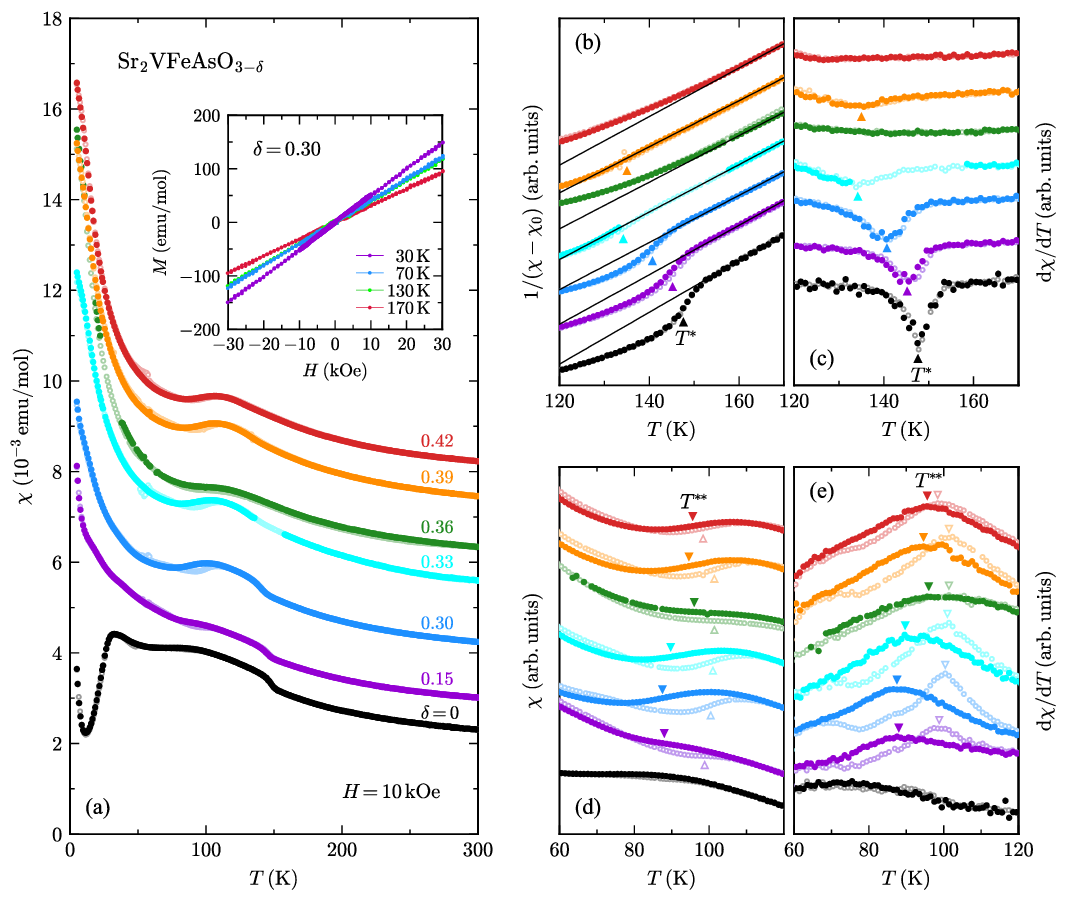}
\caption{(Color online) (a) Temperature dependence of magnetic susceptibility $\chi$ for \VFed{} measured on warming (open symbols) and cooling (closed symbols) processes under the magnetic field of 10 kOe. The data are offset for clarity. The drop of $\chi(T)$ for $\delta=0$ below $\sim 30$ K is associated with the superconducting transition. The data for $\delta=0.45$ are not shown because we observed unnaturally large $\chi$ probably due to impurity phases. Inset shows isothermal magnetization curves for $\delta = 0.30$ at various temperatures down to 30 K. (b) Temperature dependence of $1/(\chi - \chi_0)$, where $\chi_0$ is the temperature-independent term of $\chi(T)$, and (c) temperature derivative of $\chi$ from 120 K to 170 K. The solid lines represent the Curie-Weiss fitting for the cooling data above 160 K. Triangles denote \Tstar{} defined by a local minimum of $\mathrm{d}\chi/\mathrm{d}T$. (d) Magnified view of $\chi$ and (e) its temperature derivative from 60 K to 120 K. Open and closed triangles denote \Tsstar{} for warming and cooling processes, respectively, defined by a local maximum of $\mathrm{d}\chi/\mathrm{d}T$.}
\label{chi}
\end{figure*}

Figure \ref{chi}(a) shows the temperature dependence of magnetic susceptibility $\chi(T)$ of \VFed{} under $H$ = 10 kOe. $\chi(T)$ obeys the Curie-Weiss behavior at high temperatures above $\sim 150$ K: $\chi(T) = C/(T - \theta_{\mathrm{CW}}) + \chi_0$, where $C$ is the Curie constant, and $\theta_{\mathrm{CW}}$ is the Curie-Weiss temperature. The temperature-independent term $\chi_0$ likely arises from the contribution from Fe electrons, which only shows a weak temperature dependence~\cite{Johnston2010}. The Curie-Weiss fitting yielded roughly similar results for the present samples of \VFed{} with $\theta_{\mathrm{CW}} \simeq 30$ K and $\chi_0 \simeq 1.5 \times 10^{-3}$ emu/mol. The effective magnetic moment $\mu_{\mathrm{eff}}$ was estimated from $C$ to be $\sim 1.3\ \mu_{\mathrm{B}}$. 

\begin{figure}
\includegraphics[scale=1.00]{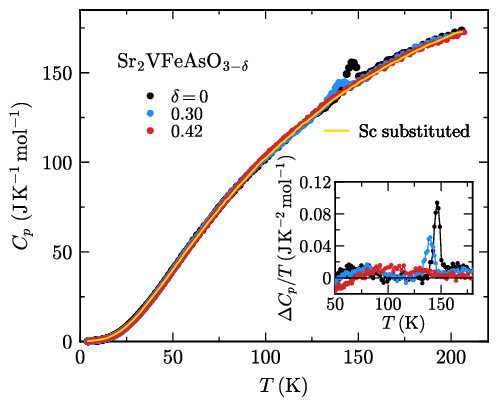}
\caption{(Color online) Temperature dependence of specific heat $C_{p}$ for $\delta = 0$, 0.30, and 0.42. As a reference, $C_{p}(T)$ for the Sc-substituted sample (nominal composition of Sr$_2$V$_{0.85}$Sc$_{0.15}$FeAsO$_3$), which shows no anomalies in $\chi(T)$ at \Tstar{} or \Tsstar{}, is displayed by the solid curve. Inset shows the temperature dependence of $\Delta C_{p} / T$, where $\Delta C_{p}$ is the specific heat after subtracting the data of the Sc-substituted sample.}
\label{Cp}
\end{figure}

With decreasing temperature, $\chi$ for low $\delta$ shows an increase at \Tstar{} $\simeq 150$ K. As shown in Fig.~\ref{chi}(b), $1/(\chi-\chi_0)$ shows a clear downward deviation from the Curie-Weiss behavior indicated as a solid line. We defined \Tstar{} as a temperature at which $\mathrm{d}\chi/\mathrm{d}T$ takes a local minimum [Fig.~\ref{chi}(c)]. The peak of specific heat $C_{p}$ at \Tstar{} evidences the presence of the phase transition (Fig.~\ref{Cp}), consistent with the previous studies~\cite{Cao2010,Tatematsu2010,Sefat2011}. With increasing $\delta$, the anomaly of $\chi(T)$ at \Tstar{} becomes smeared, along with a decrease in \Tstar{}. For the samples showing no discernible anomaly ($\delta = 0.36$ and 0.42), $1/(\chi-\chi_0)$ shows a smooth upward deviation from the Curie-Weiss behavior. The suppression of the phase transition at \Tstar{} is also confirmed by $C_{p}(T)$. As shown in the inset of Fig.~\ref{Cp}, no pronounced peak was observed in $\Delta C_{p} / T$ for $\delta=0.42$, where $\Delta C_{p}$ is the specific heat after subtracting the data of the Sc-substituted sample as the lattice contribution.

A notable feature observed in this study is a hump structure of $\chi(T)$ at $\sim 100$ K with a thermal hysteresis. As shown in Fig.~\ref{chi}(d), this behavior is most prominent for the sample with $\delta=0.30$, in which \Tsstar{}, defined as a local maximum of $\mathrm{d}\chi/\mathrm{d}T$, shows a large difference between the warming and cooling processes [Fig.~\ref{chi}(e)]. The magnetization curve, shown in the inset of Fig.~\ref{chi}(a), appears to be linear down to 30 K. Although the development of spontaneous magnetization cannot be completely ruled out, it is, if present, much smaller than observed in the previous studies~\cite{Cao2010,Tegel2010}. No clear local maximum in $\mathrm{d}\chi/\mathrm{d}T$ was observed for $\delta=0$, although a weak thermal hysteresis can be recognized in $\chi(T)$. For the samples with large $c$ ($\delta = 0.36$ and 0.42), the hump structure is disappearing, but the thermal hysteresis remains, which is clearly seen in Fig.~\ref{chi}(e). As shown in Fig.~\ref{Cp}, we observed no peak structure in $C_{p}(T)$ at \Tsstar{}, indicating that the release of entropy at \Tsstar{} is small.

\begin{figure}
\includegraphics[scale=0.94]{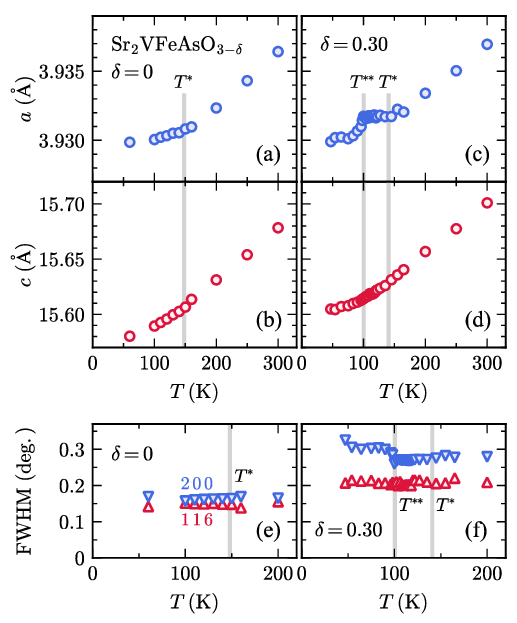}
\caption{(Color online) Temperature dependence of (a) $a$- and (b) $c$-axis lattice parameter of \VFed{} for $\delta = 0$. (c), (d) Same as (a), (b) but for $\delta = 0.30$. The data were taken on the warming process. \Tstar{} and \Tsstar{}, determined from $\chi(T)$, are indicated by the vertical gray lines. Temperature dependence of the full width at half maximum (FWHM) of the $2\hspace{0.05em}0\hspace{0.05em}0$ and $1\hspace{0.05em}1\hspace{0.05em}6$ peaks for (e) $\delta=0$ and (f) 0.30.}
\label{PF}
\end{figure}

Figures \ref{PF}(a)--\ref{PF}(d) show the temperature dependence of the lattice parameters for $\delta=0$ and 0.30. The lattice parameters were calculated from the XRD patterns assuming that the system holds the tetragonal $P4/nmm$ space group because we observed neither peak splitting nor appearance of new peaks for the measured temperature range. One can see that the phase transition at \Tstar{} makes an influence on the crystal structure, especially for $\delta=0$. A decrease in $c$ at \Tstar{} was recognized, while the temperature dependence of $a$ becomes gentle below \Tstar{}, consistent with the previous study~\cite{Saito2015}. A similar behavior, though weakened, can be seen for $\delta=0.30$.

The shrinkage of the $a$-axis length was observed for $\delta=0.30$ at $\sim 100$ K, corresponding to \Tsstar{}, while $c$ does not show a significant change. Such a behavior has been reported by Saito \textit{et al.}~\cite{Saito2015}. Although the change of the crystal system is not evident from the XRD patterns, possible orthorhombic distortion is inferred from the full width at half maximum (FWHM) of the diffraction peak. As shown in Fig.~\ref{PF}(f), the FWHM of the $2\hspace{0.05em}0\hspace{0.05em}0$ peak for $\delta=0.30$ shows an abrupt increase below \Tsstar{}. This result implies the presence of a structural phase transition, which makes the [100] and [010] directions inequivalent. This can be compatible with the thermal hysteresis observed in $\chi(T)$. Note that the present orthorhombic distortion is distinct from the kind of that frequently observed in iron-based superconductors. In the latter case, the difference of the distance between the neighboring Fe atoms along the orthogonal directions breaks four-fold rotational ($C_4$) symmetry, which leads to the [110] direction inequivalent to the [1$\bar{1}$0] direction. This is not the case for the present system because the FWHM of the $1\hspace{0.05em}1\hspace{0.05em}6$ peak shows no change across \Tsstar{} [Fig.~\ref{PF}(f)].

\subsection{NMR Measurements on \VFed{} with $\delta=0.30$}

\begin{figure}
\includegraphics[scale=1.00]{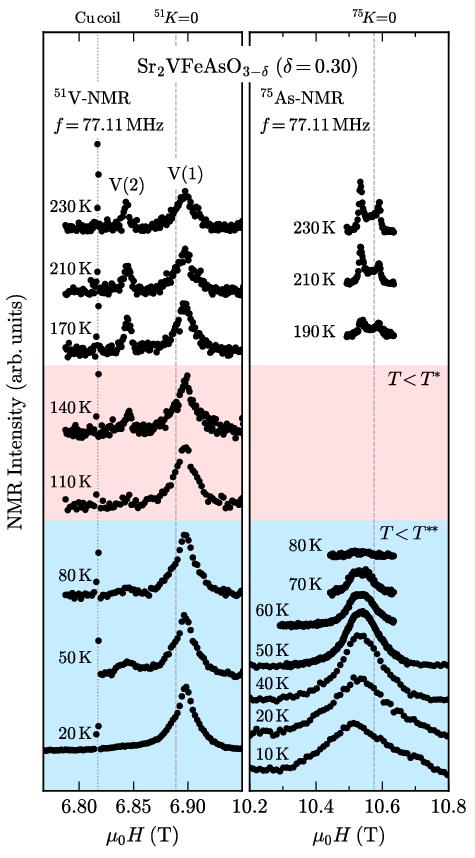}
\caption{(Color online) $^{51}$V- and $^{75}$As-NMR spectra in \VFed{} with $\delta=0.30$ for various temperatures. The vertical dashed lines display the resonance fields corresponding to zero Knight shift ($^{51/75}K=0$). The signal from a Cu coil is indicated by the vertical dotted line.}
\label{NMR}
\end{figure}

To gain insight into the contribution of V and Fe to the physical properties, we measured $^{51}$V- and $^{75}$As-NMR spectra on \VFed{} with $\delta=0.30$ [\Tc{} $\simeq 14$ K from the onset of $\chi(T)$], as shown in Fig.~\ref{NMR}. The $^{51}$V-NMR spectra are composed of two distinct peaks. The main peak, denoted as V(1), is characterized by a very weak temperature dependence both in Knight shift ($^{51}K^{(1)} \simeq -0.1\%$) and line shapes even across \Tstar{} and \Tsstar{}. This behavior is similar to the previous $^{51}$V-NMR results for \VFe{} with \Tc{} $\gtrsim 25$ K~\cite{Tatematsu2010,Kotegawa2011,Ueshima2014,Ok2017} (hereafter referred to as high-\Tc{} samples), which definitely contains much lower oxygen deficiency than \VFed{} with $\delta=0.30$ in this study. Thus, the V(1) peak can be assigned to the V site with the apical oxygen forming pyramidal VO$_5$, although the value of $^{51}K^{(1)}$ is slightly smaller than that for the high-\Tc{} samples. The smaller V(2) peak, which has not been observed for the high-\Tc{} samples so far, can be considered to arise from the V site without the apical oxygen. The observation of the two distinct V sites gives firm evidence of the presence of oxygen deficiency at the SrO layers. Note that the intensity of the V(2) site becomes weak below $\sim 100$ K and disappears at further lower temperatures, suggesting that some low-energy fluctuations of the electronic states develop locally at the V(2) sites.

The $^{75}$As-NMR spectrum at high temperatures was a typical powder pattern for the center peak, in which a two-horned shape appears due to the nuclear quadrupolar interaction. It enables us to estimate the nuclear quadrupole frequency to be $\sim 8$ MHz, which is close to the value for the high-\Tc{} sample~\cite{Kotegawa2011}. This indicates that the local electric-field gradient at the As site is not largely changed by oxygen deficiency. With decreasing temperature below 190 K, the $^{75}$As-NMR signal starts to lose its intensity as the temperature approaches \Tstar{}. This is attributed to the shortening of the spin-spin relaxation time $T_2$~\cite{Ueshima2014}. The signal is recovered below \Tsstar{} in association with a change of the spectral shape. Here, we note that the $^{75}$As-NMR signal cannot be detected for a wide temperature range between \Tstar{} and \Tsstar{}, in contrast to the case for the high-\Tc{} samples, where the signal disappears only for a very narrow temperature range around \Tstar{}~\cite{Tatematsu2010,Kotegawa2011,Ok2017}. This implies that in \VFe{}, oxygen deficiency makes $T_2$ keep short in the wider temperature range.

With further decreasing temperature, the $^{75}$As-NMR spectrum is significantly broadened below 40 K, which should be attributed to the development of magnetic ordering at the Fe sites. The broad peak suggests that the internal magnetic field at the As sites is very small and non-uniform [$\mu_0{^{75}H_{\mathrm{int}}} \le 0.1(1)$ T], which is much smaller than those of a stripe-type antiferromagnetic (AFM) state in the typical parent compounds of iron-based superconductors~\cite{Ok2017,Sakano2019}. It is also much smaller than $\mu_0{^{75}H_{\mathrm{int}}} \le 0.5$ T observed for an isostructural compound Sr$_2$ScFeAsO$_3$~\cite{Sakano2019}, where the magnetic moment at the Fe site has been evaluated to be 0.11 $\mu_{\mathrm{B}}$ by the $^{57}$Fe M\"{o}ssbauer experiment~\cite{Munevar2011}. Thus, the magnetic moment at the Fe site of the present compound may be less than 0.02 $\mu_{\mathrm{B}}$, which would be difficult to be detected by the M\"{o}ssbauer studies~\cite{Cao2010,Munevar2011,Holenstein2021}.

\section{Discussion}

\subsection{Electronic Phase Diagram of \VFed{}}

As described earlier, the $c$-axis lattice parameter is considered to be a suitable measure of the actual amount of oxygen deficiency. Although oxygen vacancies in \VFe{} have not been clarified by the neutron scattering study~\cite{Tegel2010}, our NMR result demonstrates the presence of the two distinct V sites arising from the oxygen vacancy in the apical site.

\begin{figure}
\includegraphics[scale=0.96]{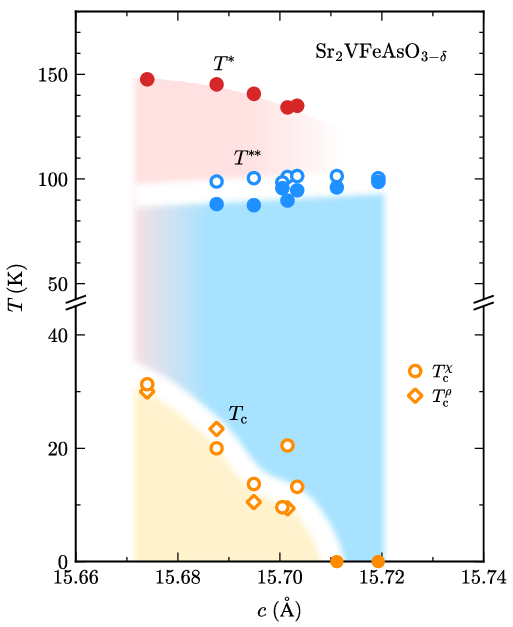}
\caption{(Color online) Electronic phase diagram of \VFed{} as a function of $c$-axis lattice parameter, which is a suitable measure of the actual amount of oxygen deficiency. \Tstar{} and \Tsstar{} are determined from the magnetic susceptibility measurement. The anomaly of $\chi(T)$ at \Tsstar{} shows different values on warming (open blue circles) and cooling (closed blue circles) processes. The plotted values of \Tc{} are determined from the onset of magnetic susceptibility (open orange circles) and zero resistivity (open orange diamonds).}
\label{PhaseDiagram}
\end{figure}

Figure \ref{PhaseDiagram} depicts the electronic phase diagram of \VFed{} summarizing the present results with the $c$-axis lattice parameter taken as a horizontal axis. The $c$-axis length becomes longer as the oxygen deficiency increases. For $\delta = 0$ with the shortest $c$-axis length in the present study, \VFed{} exhibits the phase transition at \Tstar{} = 148 K and shows superconductivity with \Tc{} exceeding 30 K. With increasing $c$, the increase in $\chi(T)$ at \Tstar{} becomes less distinct and difficult to determine unambiguously. Superconductivity is suppressed with increasing oxygen deficiency and disappears at $c \simeq 15.71$ \AA. The remarkable observation in this study is the anomaly of $\chi(T)$ at $\sim$ \Tsstar{} showing a thermal hysteresis. Such a behavior is not so appreciable for $\delta = 0$ but becomes prominent with oxygen deficiency.

Other electronic orders also appear in \VFed{}, although not represented in the phase diagram shown in Fig.~\ref{PhaseDiagram}. The present NMR result for $\delta=0.30$ suggests that the magnetic ordering of very small magnetic moment at the Fe site is present below $\sim 40$ K. This magnetic order has been confirmed for the high-\Tc{} samples by the NMR study~\cite{Ok2017} and the muon spin relaxation measurements~\cite{Munevar2011,Holenstein2021} and thus seems to be insensitive to the introduction of oxygen deficiency. The disappearance of the V(2) peak observed in the $^{51}$V-NMR spectrum (Fig.~\ref{NMR}) may correspond to the emergence of some other electronic order with a phase transition. This might be related with a finite magnetic moment of V in superconducting \VFed{} suggested by x-ray magnetic circular dichroism measurements~\cite{Horio2018}. In \VFed{}, it turned out to be possible to introduce oxygen deficiency beyond the range of the present study, and even longer $c$-axis length ($c > 15.80$ \AA) has been achieved~\cite{Tojo2019}. In such samples, large spontaneous magnetization was observed below 200--300 K, which was attributed to ferrimagnetic ordering at the V sites~\cite{Tojo2019}. The influence of these electronic orders on the phase diagram should be addressed in the future work.

\subsection{Origin of the Anomaly at \Tsstar{}}

The present study revealed that \VFed{} with oxygen deficiency exhibits the magnetic and structural anomaly at \Tsstar{} with a thermal hysteresis. We observed no significant influence on $\rho(T)$ at \Tsstar{}, implying that itinerant Fe electrons are not involved in this anomaly. Our NMR measurement suggests no magnetic ordering at around \Tsstar{}.

The broadening of the $2\hspace{0.05em}0\hspace{0.05em}0$ diffraction peak below \Tsstar{}, shown in Fig.~\ref{PF}(f), indicates the orthorhombic distortion, pointing towards the presumable orbital ordering lifting the degeneracy of the $d_{xz}$ and $d_{yz}$ orbitals of V. This is likely related with the significant influence on the $a$-axis lattice parameter across \Tsstar{} [Fig.~\ref{PF}(c)]. The present result suggests a structural phase transition, in agreement with the existence of a thermal hysteresis. It should be noted that $C_{p}(T)$ showed no anomaly at \Tsstar{}. The release of entropy would be distributed over a wide temperature range due to two dimensionality and/or a disorder effect induced by oxygen deficiency.

The present result gives some insights on the energy level of the V $t_{2g}$ orbitals. \VFe{} has a VO$_5$ pyramid unit with two $3d$ electrons per V atom. If the V atoms sit at the center of the basal plane, the energy level of the $d_{xy}$ orbital is higher than that of the $d_{xz/yz}$ orbital. However, the position of the V atoms is actually off the basal plane towards the apical oxygen~\cite{Zhu2009,Cao2010,Tegel2010,Corkett2014}, resulting in less separated or even reversed energy levels between the $d_{xy}$ and $d_{xz/yz}$ orbitals. With the energy level of the $d_{xy}$ orbital higher than that of the $d_{xz/yz}$ orbital, the lifting of the degeneracy of the $d_{xz}$ and $d_{yz}$ orbitals does not lead to an energy gain. This is the case for the V(2) sites without the apical oxygen. Thus, the structural anomaly at \Tsstar{}, indicative of the separated energy levels of the $d_{xz}$ and $d_{yz}$ orbitals, originates from the V(1) sites forming VO$_5$ pyramids with the $d_{xy}$ orbital sitting the lowest energy level.

It can be expected that fluctuations of the V orbitals exist above \Tsstar{}, leading to instability of a local structure in \VFed{}. This would influence not only the VO$_5$ pyramid but also the atomic position of As located above/below the V site [see Fig.~\ref{XRD}(a)]. The instability of the As position gives rise to a significant influence on the electronic state of Fe. The change in a relative position of As to Fe, as well as the modified electronic state of Fe, should affect the $^{75}$As-NMR spectrum. This can be a feasible explanation for the disappearance of the $^{75}$As-NMR signal for a wide temperature range in the present study. In contrast to the previous NMR studies for the high-\Tc{} samples, which demonstrated that the $^{75}$As-NMR signal recovers below \Tstar{}~\cite{Kotegawa2011,Ok2017}, the signal for the present sample with oxygen deficiency recovered below \Tsstar{}, not \Tstar{}. This implies that the $T_2$ shortening is related with the presence of V orbital fluctuations. The introduction of oxygen deficiency would produce the orbital fluctuations linked to the anomaly at \Tsstar{}, which makes the signal remain lost even below \Tstar{} and become discernible once the temperature drops below \Tsstar{}.

\subsection{Insight on the Phase Transition at \Tstar{}}

The origin of the phase transition at \Tstar{} remains unclear. There is neither magnetic ordering nor a change of the crystal system at \Tstar{}, which preserves the $C_4$ symmetry of the system. It has been debated as to whether Fe or V plays a dominant role~\cite{Tatematsu2010,Cao2010,Kotegawa2011,Ueshima2014,Ok2017}.

A plausible scenario proposed by Ok \textit{et al.} is that the frustration between fluctuations of Fe stripe-type and V N\'{e}el-type antiferromagnetism induces a charge/orbital order in the FeAs layers below \Tstar{}~\cite{Ok2017}. The NMR studies on \VFe{} demonstrated that the $^{75}$As-NMR spectrum, dominantly affected by Fe spins, exhibits a significant change at \Tstar{}, whereas the $^{51}$V-NMR spectrum remains essentially unchanged~\cite{Ok2017}. The angle-resolved photoemission spectroscopy measurement revealed the disappearance of the large hole Fermi pocket below \Tstar{}, indicative of a significant modification of a low-energy band structure arising from Fe electrons~\cite{Kim2021}. Unlike most iron-based superconductors, \VFe{} shows no stripe-type AFM ordering at the Fe sites. With decreasing temperature from room temperature, the spin-lattice relaxation rate $1/T_1$ of $^{75}$As shows an increase, indicative of an enhancement of AFM fluctuations at the Fe sites~\cite{Ueshima2014,Ok2017}. $1/T_1$ turns to decrease below $\sim 200$ K without showing a magnetic order. Probably, the presence of the perovskite-type layer hampers the lattice deformation driven by the FeAs layer, which prevents a structural phase transition necessary to the stabilization of the stripe-type AFM ordering. The suppression of the AFM spin fluctuations below 200 K would arise from the development of fluctuations of the nonmagnetic electronic order realized at \Tstar{}.

It remains possible that the V orbitals also play a key role in the phase transition at \Tstar{}. Isostructural Sr$_2$VCoAsO$_3$ exhibits a phase transition at 140 K~\cite{Ohta2018}, showing an anomaly in $\chi(T)$ similar to that observed in \VFed{}. This implies the possibility that the transition is driven by the vanadium-oxide layer. Although the contribution of V spins to the transition at \Tstar{} is very small, the previous NMR study indicated that $1/T_2$ of $^{51}$V shows a peak at \Tstar{}.~\cite{Holenstein2021} The phase transition at \Tstar{} may be accompanied by V orbital ordering without breaking the $C_4$ symmetry, giving rise to a change in the energy difference between the $d_{xz/yz}$ and $d_{xy}$ orbitals of V. Such an orbital order has been observed for $R$VO$_3$ ($R$: rare earth) at $\sim 200$ K~\cite{Miyasaka2003}, which is a temperature range close to \Tstar{} in the present system. The reduction in the $c$-axis length of \VFed{} at \Tstar{} suggests a shortening of either the As height from the Fe plane or the distance between the V atom and its upper/lower As or apical O atom. The latter leads to a relatively lower energy level of the $d_{xy}$ orbital, consistent with the discussion in the previous section. The shift of the As position results in the deformation of the FeAs$_4$ tetrahedron, which would make a pronounced impact on the $^{75}$As-NMR spectrum. As discussed in the previous section, the V orbital fluctuations associated with the anomaly at \Tsstar{} seem to be influential on the undetected $^{75}$As-NMR signal for \VFed{} with oxygen deficiency. In this context, the disappearance of the $^{75}$As-NMR signal for the high-\Tc{} samples around \Tstar{}~\cite{Kotegawa2011,Ok2017} may also arise from the orbital fluctuations associated with the phase transition at \Tstar{}. For the high-\Tc{} samples (e.g., $\delta=0$ in this study), orbital fluctuations would strongly diminish after the phase transition at \Tstar{} because the very faint anomaly at \Tsstar{} indicates the absence of substantial orbital fluctuations at temperatures below \Tstar{}.

\subsection{Effect of Oxygen Deficiency on Superconductivity}

Superconductivity in \VFed{} tends to be suppressed with oxygen deficiency, as can be seen in the phase diagram shown in Fig.~\ref{PhaseDiagram}. This tendency coincides with the previous studies~\cite{Zhu2009,Tatematsu2010,Cao2010,Tegel2010,Sefat2011,Ok2018,Wakimura2019,Tojo2019,Iwasaki2020}, although the already reported values of \Tc{} are rather scattered. The $c$-axis length at which superconductivity disappears is close to that the anomaly of $\chi(T)$ at \Tstar{} becomes indiscernible. The emergence of superconductivity in this system might be related with the presence of the transition at \Tstar{}, as pointed out by the previous study~\cite{Ok2018}.

The effect of disorder is likely to be a leading factor that suppresses superconductivity. As can be seen in Fig.~\ref{resistivity}, the reduction of the residual resistivity ratio as well as the low-temperature upturn was observed. This indicates that \VFed{} is a strongly disordered system, although oxygen deficiency is considered to be introduced in the SrO layers, which is located away from the FeAs layers responsible for superconductivity. Such a disorder effect seems to be larger than that in Co-doped \VFe{}, in which chemical substitution takes place in the conducting FeAs layers.~\cite{Ok2018} For other iron-based superconductors, no significant increase of residual resistivity has been observed with chemical substitution~\cite{Shekhar2008,Nakajima2014} or oxygen deficiency~\cite{Ishida2010} away from the FeAs layers. In the present system, the disorder effect would be very large due to strong two dimensionality.

Another possible factor is the variation in the local structure. It has been known that iron-based superconductors show high \Tc{} when the As-Fe-As bond angle is close to 109.5$^\circ$ forming a regular FeAs$_4$ tetrahedron~\cite{Lee2012}. The bond angle of \VFe{} showing \Tc{} $>$ 30 K is close to this value~\cite{Tegel2010,Ok2018}. It has been reported that isovalent substitution of Ca for Sr~\cite{Corkett2014} and application of physical pressure~\cite{Garbarino2011} deform the FeAs$_4$ tetrahedron away from the optimal shape, resulting in the suppression of superconductivity. The introduction of oxygen deficiency may suppress superconductivity via the variation in the local crystal structure, although strictly, a change in structure at \Tstar{} and \Tsstar{} should be taken into account.

In the present system, electron doping due to oxygen deficiency does not seem to be a dominant factor of the \Tc{} variation. The electrons introduced by oxygen deficiency would be distributed in the vanadium-oxide and/or FeAs layers. The change in Hall coefficient $R_\mathrm{H}$ suggests that at least a part of electrons are doped into the FeAs layers, but the value of $R_\mathrm{H}$ shows no correlation with \Tc{} (see Appendix). Although it has been reported that \Tc{} of \VFe{} decreases with Co doping~\cite{Ok2018} and Cr doping~\cite{Wakimura2019}, corresponding to electron doping, the carrier doping effect is likely to be less important in \VFed{} than the disorder effect and the variation in the local structure.

\section{Summary}

We systematically investigated the electronic state of \VFed{} with various amount of oxygen deficiency $\delta$ and demonstrated the phase diagram of \VFed{} as a function of the $c$-axis lattice parameter. As oxygen deficiency is introduced, the $c$-axis length is elongated. With increasing $c$, \Tc{} as well as \Tstar{} decreases and disappears at $c \simeq 15.71$ \AA. It was found that the magnetic and structural anomaly with a thermal hysteresis prominently shows up at \Tsstar{} for the samples with oxygen deficiency. The low-temperature XRD measurement revealed the orthorhombic distortion below \Tsstar{}, suggesting the presence of the V orbital ordering with the energy difference between the $d_{xz}$ and $d_{yz}$ orbitals. The $^{75}$As-NMR signal of \VFed{} with $\delta=0.30$ cannot be detected for a wide temperature range between \Tstar{} and \Tsstar{}. These features probably arise from the instability of the As position induced by the fluctuations of the V orbitals. The change in the atomic position of As would significantly affect the electronic state of the FeAs layer. Our findings demonstrate that besides the dominant contribution of the FeAs layer to electrical conduction, the insulating vanadium-oxide layer plays a significant role in the electronic state of \VFed{}.


\begin{acknowledgment}


We would like to thank N. Murai for fruitful discussions. This work was supported by JSPS KAKENHI (Grant Numbers JP18K13500 and JP22H04479), the Toyota Riken Scholar Program, Nippon Sheet Glass Foundation for Materials Science and Engineering, Izumi Science and Technology Foundation, the Casio Science Promotion Foundation, and Takahashi Industrial and Economic Research Foundation. The XRD experiments at PF were carried out under the approval of the Photon Factory Program Advisory Committee (Proposal No. 2016G145).

\end{acknowledgment}

\appendix
\section{Hall Effect}

\begin{figure}
\includegraphics[scale=0.96]{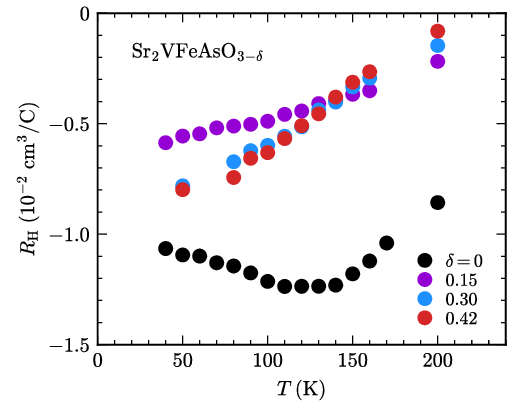}
\caption{(Color online) Temperature dependence of Hall coefficient $R_\mathrm{H}$ of \VFed{} for $\delta = 0$, 0.15, 0.30, and 0.42.}
\label{Hall}
\end{figure}

Figure \ref{Hall} shows the temperature dependence of Hall coefficient $R_\mathrm{H}$ for \VFed{} with $\delta=0$, 0.15, 0.30, and 0.42. The negative Hall coefficient indicates the dominant contribution of electron carriers to electrical conduction. A strong temperature dependence of $R_\mathrm{H}$ arises from an effect of band multiplicity typical of iron-based superconductors. With increasing $\delta$, the absolute value of $R_\mathrm{H}$ tends to decrease, as reported in the previous studies~\cite{Han2010,Sefat2011}. This tendency is consistent with electron doping induced by oxygen deficiency, but the variation of $R_\mathrm{H}$ shows no clear correlation with the amount of oxygen deficiency and hence \Tc{}. For $\delta=0$, one can clearly see a change in temperature dependence across \Tstar{}. This change is rapidly suppressed with oxygen deficiency. No discernible change at \Tsstar{} was observed in the present study.

\bibliographystyle{jpsj}
\bibliography{21113Odef}


\end{document}